\documentclass[aps,english,twocolumn,showpacs]{revtex4}
\usepackage[T1]{fontenc}
\usepackage[latin9]{inputenc}
\usepackage{color}
\usepackage{babel}

\usepackage{amsmath}
\usepackage[unicode=true, 
 bookmarks=true,bookmarksnumbered=false,bookmarksopen=false,
 breaklinks=false,pdfborder={0 0 1},backref=false,colorlinks=true]
 {hyperref}
\hypersetup{pdftitle={A k-essence Model of Inflation, Dark matter and Dark Energy},
 pdfauthor={Nilok Bose, A. S. Majumdar},}



\begin{document}

\title{A \textit{k}-essence Model Of Inflation, Dark Matter and Dark Energy }

\author{Nilok Bose}

\email{nilok@bose.res.in}

\author{A. S. Majumdar}

\email{archan@bose.res.in}

\affiliation{S. N. Bose National Centre for Basic Sciences, Block JD, Sector III,
Salt Lake, Calcutta 700098, India}

\date{\today}
\begin{abstract}
We investigate the possibility for \textit{k}-essence dynamics to
reproduce the primary features of inflation in the early universe,
generate dark matter subsequently, and finally account for the presently
observed acceleration. We first show that for a purely 
kinetic \textit{k}-essence
model the late time energy density of the universe when expressed
simply as a sum of a cosmological constant and a dark matter term
leads to a static universe.
We then study another \textit{k}-essence model in which the Lagrangian
contains a potential for the scalar field as well as a non-canonical
kinetic term. We show that such a model generates the basic features
of inflation in the early universe, and also gives rise to dark matter
and dark energy at appropriate subsequent stages. Observational constraints
on the parameters of this model are obtained. 
\end{abstract}

%\keywords{Inflation, dark matter, dark energy, \textit{k}-essence}

\pacs{98.80.-k, 98.80.Cq, 95.36.+x}

\maketitle

\section{Introduction}

Over the course of the past decade, evidence for the most striking
result in modern cosmology has been steadily growing, namely the current
acceleration of the universe. The nature of the physical mechanism
driving this acceleration is yet unclear, though there exists an increasingly
wide variety of approaches that could theoretically account for the
present acceleration. The proposal of a cosmological constant to generate
the late time acceleration of the universe is consistent with several
important observations such as the red-shift of distant supernovae,
the power spectrum of the cosmic microwave background (CMB), and the
distribution of large scale structure. However, there is no compelling
theoretical explanation for its actual value that could account for
the cosmic coincidence problem as to why the accelerating phase should
have begun in a narrow window of time in the present universe. Nonetheless,
observations have categorized the energy density of the present universe
to consist of approximately 23\% dark matter, which clusters and drives
the formation of large-scale structure in the universe, and 73\% dark
energy, which drives the late-time acceleration of the universe (See
\cite{wmap}, \cite{sahni}, \cite{sami} and references therein).

Since the nature of both dark matter and dark energy are unknown,
it is plausible that these two mysterious components of the universe
are the manifestations of a single entity. Several examples of attempts
to unify dark matter and dark energy can be found in the literature
(for instance \cite{padma}, \cite{scherrer}, \cite{arbey}). Further,
it is very strongly believed that there was an early inflationary
period of the universe, and the nearly scale independent density perturbations
produced during inflation have left a faithful imprint on features
of the CMB power spectrum. The idea that a single or similar mechanism
could be responsible for accelerating evolutions of the universe both
at early and late times has received the attention of physicists with
models constructed to explain inflation and dark energy using a single
scalar field (for eg. quintessential inflation \cite{peebles}). Apart
from the above category of models, schemes that try to unify dark
matter and inflation can be found in the literature (for instance
\cite{lidsey}). Also there have been attempts to unify inflation,
dark matter and dark energy (for instance \cite{triple}). In this
paper our motivation is to explore a possibility for achieving a triple
unification, \textit{viz.} inflation, dark matter and dark energy
within the context of the same model. With this aim we investigate
the dynamics of a \textit{k}-essence scalar field model.

The idea of \textit{k}-essence motivated from the Born-Infeld action
of string theory \cite{born-infeld}, was first introduced as a possible
model for inflation \cite{picon}, \cite{garriga}. Later, it was
noted that \textit{k}-essence could also yield interesting models
for the dark energy \cite{chiba}, \cite{picon2}, \cite{picon3},
\cite{chiba2}, \cite{chimento}, \cite{chimento2}. A parallel mechanism
for producing the late time acceleration of the universe through the
dynamics of scalar fields, \textit{viz.} quintessence \cite{quint},
has also gained a lot of popularity in the literature. In most of
the quintessence models the late time dynamics is dominated by the
potential for the scalar field. A crucial difference between quintessence
and \textit{k}-essence is that the latter class of models contain
non-canonical kinetic terms in the Lagrangian. In this sense quintessence
may also be viewed as a special case of \textit{k}-essence. Another
important subset of \textit{k}-essence is purely kinetic \textit{k}-essence
in which the Lagrangian contains only a kinetic factor, i.e., a function
of the derivatives of the scalar field, and does not depend explicitly
on the field itself. Such models were in fact, the first ones investigated
in the context of inflation \cite{picon}. In this context, they successfully
yield exponential inflation, but suffer from the ``graceful exit''problem.

An interesting attempt was made to unify dark matter and dark energy
using kinetic \textit{k}-essence in \cite{scherrer}. Though this
model had its share of problems as pointed out by the author, extensions
of the formalism to extract out dark matter and dark energy components
within a unified framework have been used also in subsequent works
\cite{chimento3}. It is worth noting that a purely 
kinetic \textit{k}-essence leads to a static universe when the late time 
energy density of the universe is expressed simply as
a sum of a cosmological constant and a dark matter term.
In the present paper we first provide an argument in support of this
fact. This sets the stage for study of our model which contains both
a potential and a non-canonical kinetic term for the scalar field,
but where it is possible to use a part of the formalism of \cite{scherrer},
as we show subsequently. We then develop our \textit{k}-essence model
that causes inflation in the early universe and behaves as purely
kinetic \textit{k}-essence in the late universe and reproduces a cosmological
constant and a dark matter term in the energy density. In order to
discuss the viability of our model, we further provide estimates of
the the values of the model parameters that could be obtained from
observational constraints.

\section{Purely Kinetic \textit{k}-essence \label{sec:2}}

A general \textit{k}-essence Lagrangian can be written as \begin{equation}
{\cal L}\,\,\equiv\,\,{\cal L}\,\left(\phi,\, X\right)\label{1}\end{equation}
where $\phi$ is the scalar field and \textit{X} is defined as \begin{equation}
X\,=\,\frac{1}{2}\,\partial_{\mu}\,\phi\,\partial^{\mu}\,\phi\label{2}\end{equation}
The Lagrangian can be any function of the scalar field and \textit{X}.
In this work we will consider a Lagrangian of the type \begin{equation}
{\cal L}\,\,=\,\, F\left(X\right)\,-\, V\left(\phi\right)\label{3}\end{equation}

Such a form has previously been studied in the context of \textit{k}-essence
models in \cite{sami}, \cite{mukhanov},\cite{vikman}, \cite{babichev}
and its properties have been discussed in some detail in \cite{fang}.
Note though that another class of models of the type ${\mathcal{L}}=F(X)V(\phi)$
have also been widely used in the literature \cite{chiba}, \cite{picon2},
\cite{picon3}, \cite{chiba2}, \cite{chimento}, \cite{chimento2}.

For purely kinetic k-essence one has \begin{equation}
{\cal L}\,=\,\, F\left(X\right)\label{4}\end{equation}
The energy density in these models is given by \begin{equation}
\rho\,\,=\,\,2\, X\, F_{X}\,-\, F\label{5}\end{equation}
where $F_{X}\,=\, dF/dX$ , and the pressure \textit{p} is simply
given by Eq.\eqref{4}. Therefore, the equation of state parameter
$w\,\,\equiv\,\, p/\rho$ is \begin{equation}
w\,\,=\,\,\frac{F}{2\, X\, F_{X}-\, F}\label{6}\end{equation}
The adiabatic sound speed for such models , following the convention
of Ref. \cite{garriga}, is defined to be \begin{equation}
c_{s}^{2}\,\,\equiv\,\,\frac{\partial p/\partial X}{\partial\rho/\partial X}\,\,=\,\,\frac{F_{X}}{2\, X\, F_{XX}\,+\, F_{X}}\label{7}\end{equation}
where $F_{XX}\,=\, d^{2}F/dX^{2}$ .

Throughout this paper we work in a flat Robertson-Walker metric. Now,
the equation for the kinetic \textit{k}-essence field is \begin{equation}
\left[F_{X}\,\,+\,\,2X\, F_{XX}\right]\ddot{\phi}\,\,+\,\,3HF_{X}\,\dot{\phi}=\,\,0\label{8}\end{equation}
which if rewritten in terms of \textit{X} turns out to be \begin{equation}
\left[F_{X}\,\,+\,\,2X\, F_{XX}\right]\dot{X}\,+\,\,6HF_{X}\, X\,\,\,\,=\,\,0\label{9}\end{equation}
This can be integrated exactly \cite{scherrer}, to give the solution
\begin{equation}
\sqrt{X}\, F_{X}\,\,=\,\, ka^{-3}\label{10}\end{equation}
where \textit{k} is a constant of integration. This solution was previously
derived in a slightly different form in Ref. \cite{chimento2}. Given
any form of \textit{F}(\textit{X}) equation \eqref{10} gives the
evolution of \textit{X} as a function of $a$. This result holds irrespective
of the spatial curvature of the universe.

Let us now see if the pure kinetic \textit{k}-essence model could
account for both dark matter and dark energy in a straightforward manner. 
As the most simple choice for the configuration of the late time energy
density, let us express the \textit{k}-essence energy density as 
\begin{equation}
\rho\,\,=\,\,\lambda\,\,+\,\,\frac{C_{1}}{a^{3}}\label{11}\end{equation}
where the energy density is the sum of a cosmological constant and
matter-like term which we call dark matter. Needless to say, such an
expression will hold as the true energy density of the universe after
matter domination has begun, i.e., when radiation is a negligible
fraction of the total energy density of the universe. Now using \eqref{10}
we can rewrite the energy density in \eqref{11} in terms of $X$ as \begin{equation}
\rho\,\,=\,\,\lambda\,\,+\,\,\frac{C_{1}}{k}\,\sqrt{X}\, F_{X}\label{12}\end{equation}
Equating \eqref{5}) (which gives the expression for the standard form of the
energy density for a purely kinetic \textit{k}-essence model) with \eqref{12} we get a differential equation
for \textit{F} given by \begin{equation}
\frac{F_{X}}{\lambda\,\,+\,\, F}\,\,=\,\,\frac{1}{2\, X\,\,-\,\,\dfrac{C_{1}}{k}\,\sqrt{X}}\label{13}\end{equation}
On integrating this equation one gets \begin{equation}
F\,\,=\,\,-\lambda\,\,-C_{2}\left(C_{1}\,\,-\,\,2k\sqrt{\, X}\right)\label{14}\end{equation}
with $C_{2}$ being an integration constant. Note here that since $x$ and
$a$ are related by Eq.(10), the constancy of $C_2$ with respect to $x$ implies
constancy of $C_2$ with respect to $a$ as well. 
Now, using the relation \eqref{10}
once again to switch back to the variable $a$ in the expression for $F$ in
Eq.\eqref{14}, one obtains \begin{equation}
C_{2}\,\,\,=\,\,\frac{1}{a^{3}}\label{15}\end{equation}
Thus, the only solution compatible with the ansatz (Eq.\eqref{11}) for 
the energy
density is of a constant scale factor $a$. Such a solution is indeed
consistent with the specific form for $F(X)$ in Eq.\eqref{14} (actually follows
from it). However, since this solution is not compatible with an observationally
expanding universe,
it rules out our assumption of the energy density to be of the
form expressed in Eq. \eqref{11}. We must clarify here that we have
assumed that the kinetic \textit{k}-essence energy density to be exactly
of the form of \eqref{11}, whereas in \cite{scherrer}, \cite{chimento3}
the resultant energy density came out to be approximately of the form
of \eqref{11} under certain assumptions. Therefore, using purely kinetic
\textit{k}-essence we cannot hope to unify dark matter \& dark energy,
at least exactly in the form of \eqref{11}. Nonetheless, our analysis does
not rule out other possible functional categorizations of the late time
energy density through which dark matter and dark energy could possibly
emerge.  One could also look into
other avenues to achieve the unification, and we try to provide such a
way with our model in the next section.

\section{ The Model}

We choose our Lagrangian of the form \begin{equation}
\,{\cal L}\,=\,\, F\left(X\right)\,\,-\,\, V\left(\phi\right)\label{16}\end{equation}
As stated earlier such forms have previously appeared within the context
of \textit{k}-essence models in Refs. \cite{sami}, \cite{mukhanov},
\cite{vikman}, \cite{babichev} and \cite{fang}. Several functional
forms of $F$ have been used earlier in the literature. Here we work
with a somewhat general form \begin{equation}
F\left(X\right)\,\,=\,\, K\, X\,\,-\,\, m_{Pl}^{2}\,\, L\,\sqrt{X}\,\,+\,\, m_{Pl}^{4}\,\, M\label{17}\end{equation}
where $K$, $L$ and $M$ are dimensionless positive constants, and
keeping with the spirit of \textit{k}-essence, the second term represents
the non-canonical correction ($L^{2}>4KM$) to the kinetic energy.
Our choice of the form of $F\left(X\right)$ is similar to the type
considered in Ref. \cite{chimento2}. Additionally, we include a nonvanishing
potential $V(\phi)$ given by \begin{equation}
V\left(\phi\right)=\frac{1}{2}m^{2}\phi^{2}\label{18}\end{equation}
In order to make the subsequent analysis more transparent, especially
while applying observational constraints on the parameters, let us
rewrite the kinetic part of our Lagrangian in the form \begin{equation}
F(X)\,\,=\,\, B\left(1\,\,-\,\,2\, A\,\sqrt{X}\right)^{2}-\,\, C\label{19}\end{equation}
where $A$, $B$, and $C$ can be expressed in terms of our original
model parameters as \begin{equation}
A=m_{Pl}^{-2}\frac{K}{L};\,\, B=m_{Pl}^{4}\frac{L^{2}}{4K};\,\, C=m_{Pl}^{4}(\frac{L^{2}}{4K}-M)\label{20}\end{equation}
The energy density corresponding to our model turns out to be \begin{equation}
\begin{array}{l}
{\rho\,\,\,=\,\,\,2X\, F_{X\mathrm{\;}}-\,\, F\,\,+\,\, V}\\
\\{\mathrm{\;\;}\,\,=\,\,\, B\left(4A^{2}X-1\right)\,\,\,+\,\,\, C\,\,\,+\,\,\,\frac{1}{2}m^{2}}\phi^{2}\end{array}\label{21}\end{equation}
and the pressure is given by \begin{equation}
p\ =\ B\left(1-2A\sqrt{X}\right)^{2}\ -\ C\ -\ \frac{1}{2}m^{2}\phi^{2}\label{22}\end{equation}
The Friedmann equation in this case can be written as \begin{equation}
H^{2}=\frac{8\pi G}{3}\left(4A^{2}BX-B+C+\frac{1}{2}m^{2}\phi^{2}\right)\label{23}\end{equation}
The equation of motion for the scalar field is obtained to be \begin{equation}
\left[F_{X}\,\,+\,\,2X\, F_{XX}\right]\ddot{\phi}\,\,+\,\,3HF_{X}\,\dot{\phi}\,\,\,\,+\frac{dV}{d\phi}=\,\,0\label{24}\end{equation}
which in terms of the parameters can be written as \begin{equation}
4A^{2}B\ddot{\phi}\mathrm{\;}+\mathrm{\;\;}12HA^{2}B\,\dot{\phi}{\mathrm{\;\;}-\mathrm{\;\;}6\sqrt{2}HAB+\mathrm{\;\;}m^{2}\phi\mathrm{\;\;}=\mathrm{\;\;}0}\label{25}\end{equation}
Considering the standard slow-roll approximation for inflation we
initially take the potential to be much larger than the kinetic part
, i.e. we have $V\left(\phi\right)>>2XF_{X}-F$. Correspondingly the
field equation \eqref{24} approximates to \begin{equation}
3HF_{X}\,\dot{\phi}\,\,\,\,+\frac{dV}{d\phi}\simeq\,\,0\label{26}\end{equation}
and we can write Eq.\eqref{23} as \begin{equation}
H^{2}\simeq\frac{8\pi G}{3}\left(\frac{1}{2}m^{2}\phi^{2}\right)\label{27}\end{equation}
The slow-roll parameters for this model are given by ($V^{\prime}\,=\, dV/d\phi$
\ \& $V^{\prime\prime}\ =\ d^{2}V/d\phi^{2}$) \begin{equation}
\epsilon\,\,=\frac{1}{16\pi G}\left(\frac{V^{\prime}}{V}\right)^{2}\frac{1}{F_{X}}\label{28}\end{equation}

\begin{equation}
\eta\,\,=\,\,\frac{1}{8\pi G}\frac{V^{\prime\prime}}{V}\frac{1}{F_{X}^{2}}\label{29}\end{equation}
It can be seen that the slow-roll parameters for this model are similar
to the standard inflationary scenario, the only difference being the
extra factors of $F_{X}$. To completely identify with the standard
case we demand that $F_{X}\,\,\sim\,\, O\left(1\right)$. Now equating
Eqs.\eqref{26} and \eqref{27} one obtains \begin{equation}
\sqrt{X}\,\,\simeq\,\,\frac{1}{4\sqrt{2}\, A^{2}\, B}\,\left(-\,\,\frac{m}{\sqrt{12\pi G}}\,\,+\,\,2\sqrt{2}\, A\, B\right)\label{30}\end{equation}
showing that for the duration of inflation \textit{X} and hence $F_{X}$
are practically constant. The number of e-folds of expansion is given
by \begin{equation}
\begin{array}{l}
{N\,\,=\,\,\int_{t_{i}}^{t_{e}}H\, dt\,\,\,=\,\,\,\,8\pi G\,\int_{\phi_{e}}^{\phi_{i}}\dfrac{V}{V^{\prime}}\, F_{X}\, d\phi\,\,}\\
\\{\,\,\,\,\,\,\,\,\,\simeq\,\,4\pi G\, F_{X}\,\,\dfrac{\phi_{i}^{2}\,\,-\,\,\phi_{e}^{2}}{2}\,\,}\\
\\{\,\,\,\,\,\,\,\,\,=\,\,\dfrac{4\pi G\, F_{X}}{m^{2}}\,\left(V_{i}\,\,-\,\, V_{e}\right)}\end{array}\label{31}\end{equation}
where the subscript `\textit{i}' refers to beginning of inflation
and `\textit{e}' refers to the end. Inflation ends with $\epsilon\,\,\sim\,\,1$
leading to \begin{equation}
\phi_{e}^{2}\,\,\simeq\,\,\frac{1}{4\pi G\, F_{X}}\label{32}\end{equation}
Using this in Eq.\eqref{31} we get \begin{equation}
V_{i}\,\,\simeq\,\,\frac{m^{2}}{4\pi G\, F_{X}}\,\left(N\,\,+\,\frac{1}{2}\,\right)\label{33}\end{equation}
So far the inflationary scenario in our model is almost indistinguishable
from a standard scalar field inflation involving a chaotic quadratic
potential. As inflation ends there will be kinetic domination since
now the potential decays and becomes gradually negligible. So for
the period of kinetic domination, Eq.\eqref{24} can be approximated
as \begin{equation}
\left[F_{X}\,\,+\,\,2X\, F_{XX}\right]\ddot{\phi}\,\,+\,\,3HF_{X}\,\dot{\phi}\,\,\,\simeq\,\,0\label{34}\end{equation}
i.e., we effectively recover Eq.\eqref{8} for kinetic \textit{k}-essence.
So the formalism described in section \ref{sec:2} carries over.
Hence, using Eq.\eqref{10} we get \begin{equation}
X\,\,=\,\,\frac{1}{16\, A^{4}\, B^{2}}\left(2\, A\, B\,\,+\,\,\frac{k}{a^{3}}\right)^{2}\label{35}\end{equation}
Then using Eq.\eqref{21}, and keeping in mind that \textit{V} is
now negligible, the energy density at this stage is given by \begin{equation}
\rho\,\,=\,\, C\,\,+\,\,\frac{k}{A\, a^{3}}\,\,+\,\,\frac{k^{2}}{4A^{2}\, B\, a^{6}}\label{36}\end{equation}

The subsequent evolution of the universe may be described as follows.
During the initial period of kinetic domination the third term in
Eq.\eqref{36} dominates. But that term becomes small quickly (compared
to the radiation term $\sim a^{-4}$ that we have not written down
explicitly here) and a period of radiation domination in the universe
ensues. The second term in Eq.\eqref{36} gains prominence in the
epoch of matter domination, and we identify it with dark matter. But
as the universe evolves toward the present era the first term begins
to dominate and behaves as a cosmological constant giving rise to
the observed accelerated expansion of the universe. The equation of
state parameter after the end of inflation is \begin{equation}
w\,\,=\,\,\frac{\dfrac{k^{2}}{4A^{2}Ba^{6}}\,\,-\,\, C}{C\,\,+\,\,\dfrac{k}{A\, a^{3}}\,\,+\,\,\dfrac{k^{2}}{4A^{2}\, B\, a^{6}}}\label{37}\end{equation}
We outline the values of \textit{w} over the various epochs , which
further supports the above statements : 

\begin{flushleft}
$w\approx1$\ \ \ \ \ \ \ \ ~~~~~~ after the end of
inflation and \\
 \ \ \ \ \ \ \ \ \ \ \ \ ~~~~~~~~\ \ \ before
radiation domination 
\par\end{flushleft}

\noindent \begin{flushleft}
$w\approx0$ \ \ \ \ \ \ \ ~~~~~~ during matter domination 
\par\end{flushleft}

\noindent \begin{flushleft}
$w\rightarrow-1$ \ \ \ \ \ \ ~~~~ as $a\mathrm{\;}\rightarrow\mathrm{\;}\infty$ 
\par\end{flushleft}

Using Eq.\eqref{7} the adiabatic sound speed turns out to be \begin{equation}
c_{s}^{2}\,\,=\,\,\frac{1}{\dfrac{2A\, B\, a^{3}}{\, k}\,\,+\,\,1}\label{38}\end{equation}
From Eq.\eqref{38} we see that the sound speed gradually becomes
zero as the universe expands. In the next section we will show that
it is negligible during the era of matter domination and beyond.

\section{Observational Constraints}

We have so far seen that the model considered by us reproduces the
primary features of early inflation and gives rise to a matter as
well as a dark energy component in the later evolution of the universe.
The viability of this model depends of course on the possible values
of the parameters. Let us now use various observational features of
the universe to constrain the parameters of our model. We first discuss
the inflationary dynamics of the early universe.

The amplitude of density perturbations is given by \begin{equation}
\begin{array}{c}
{\delta_{H}\simeq\dfrac{H^{2}}{4\pi^{3/2}\dot{\phi}}=4\sqrt{\dfrac{2}{3}}G^{3/2}\dfrac{V^{3/2}}{V^{\prime}}F_{X}}\\
\\{\;\;\;\;\;\;\;\;\;\;\;\;\;\;\;\;\;\;=\dfrac{4}{\sqrt{3}}G^{3/2}mF_{X}\phi^{2}}\end{array}\label{39}\end{equation}
According to the COBE normalization $\delta_{H}\;\sim\;\mathrm{\;}2\mathrm{\;}\times\mathrm{\;}10^{-5}\mathrm{\;}$.
We assume that 60 e-folds of inflationary expansion takes place. From
Eq.\eqref{33} this then gives \begin{equation}
\phi_{i}^{2}\, F_{X}\,\,\simeq\,\,\frac{60.5}{2\pi G}\label{40}\end{equation}
Hence using this value in Eq.\eqref{39} we get $m\sim10^{13}\, GeV\,\,=\,\,10^{-6}\, m_{Pl}$.
Again using Eq.\eqref{40} we find the slow-roll parameters from Eqs.\eqref{28},
and \eqref{29} to be

\begin{equation}
\epsilon\left(\phi_{i}\right)=\frac{1}{16\pi G}\frac{4}{\phi_{i}^{2}F_{X}}=\frac{1}{2\left(N+1/2\right)}=7.63\times10^{-3}\label{41}\end{equation}

\begin{equation}
\eta\left(\phi_{i}\right)=\frac{1}{8\pi G}\frac{2}{\phi_{i}^{2}F_{X}^{2}}=\frac{1}{2\left(N+1/2\right)F_{X}}\sim O\left(10^{-3}\right)\label{42)}\end{equation}
The tensor-to-scalar ratio turns out to be \begin{equation}
r=16\ \epsilon\left(\phi_{i}\right)=0.12\label{43}\end{equation}
Similarly, the spectral index is obtained as \begin{equation}
n_{s}\,=\,\,1\,\,-\,\,6\epsilon\left(\phi_{i}\right)\,\,+\,\,2\eta\left(\phi_{i}\right)\,\,\,\approx0.95\label{44)}\end{equation}
Furthermore from Eq.\eqref{33} we see that the initial value of the
potential is \begin{equation}
V_{i}\,\approx\,\,10^{65}\,(GeV)^{4}\,\,\ll\,\, m_{Pl}^{4}\,\,=\,\,10^{76}\,(GeV)^{4}\label{45}\end{equation}
showing that classical physics remains valid at the beginning of inflation.

All the above calculated parameters are of the same magnitude as one
would get in a standard model of inflation based on a quadratic chaotic
potential. Knowing the value of \textit{m} we can also estimate the
magnitude of the kinetic component during inflation, from Eq.\eqref{30}
to be \begin{equation}
X\,\,=\,\,\frac{1}{2}\,\dot{\phi}^{2}\,\,\approx\,\,10^{62}\,(GeV)^{4}\label{46}\end{equation}
We could estimate the above value because we had assumed that $F_{X}\,\,\sim\,\, O\left(1\right)$.
In view of Eq.\eqref{46} and also since $F_{X}\,\,=\,\,4A^{2}B\,\,\,-\,\,\,\dfrac{2AB}{\sqrt{X}}$,
this assumption leads to \begin{equation}
K\,\,=\,\,4A^{2}B\,\,\sim\,\, O\left(1\right)\label{47}\end{equation}
where we have used Eq.\eqref{20} in the first equality. Now when
inflation ends then using Eq.\eqref{32} and the value of \textit{m}
we see that \begin{equation}
V_{e}\,\,=\,\,\frac{m^{2}}{8\pi GF_{X}}\,\,\approx\,\,10^{62}\, GeV^{4}\,\,\approx\,\, X\label{48}\end{equation}

Thereafter the magnitude of the potential decreases and the kinetic
component begins to dominate, and as we saw from Eq.\eqref{36} when
there is full kinetic domination it will fall of as $a^{-6}$, quickly
paving the way for a radiation dominated universe. After the end of
inflation the field $\phi$ continues to roll down in the absence
of any minimum in the potential. Thus reheating can take place only
through gravitational particle production. Standard calculations \cite{ford},
\cite{spokoiny} give the density of particles produced at the end
of inflation as \begin{equation}
\begin{array}{c}
{\rho_{R}\,\,\sim\,\,0.01\, g\, H_{e}^{4}\,\,=\,\,0.01\, g\,\left(\dfrac{8\pi G}{3}\, V_{e}\right)^{2}}\\
\\{\;\;\;\;\;\;\;\;\;\;\;\;\;\;\;\;\;\;\;\;\;\;\;\;=\,\,0.01\, g\,\left(\dfrac{m^{2}}{3F_{X}}\right)^{2}}\end{array}\label{49}\end{equation}
where \textit{g} is the number of fields which produce particles at
this stage, likely to be between 10 and 100. The relative densities
turn out to be \begin{equation}
\begin{array}{l}
{\dfrac{\rho_{R}}{\rho_{\phi}}\,\,\,=\,\,\,0.01\, g\,\left(\dfrac{m^{2}}{3F_{X}}\right)^{2}\,\dfrac{8\pi GF_{X}}{m^{2}}}\\
\\{\,\,\,\,\,\,\,\,\,\,\,\,\,=\,\,7.71\times10^{-14}\,\dfrac{g}{F_{X}}}\end{array}\label{50}\end{equation}
The numerical value of the radiation density from Eq.\eqref{49} is
\begin{equation}
\rho_{R}\,\,\simeq\,\,8.46\times10^{49}\,\,\frac{g}{F_{X}^{2}}\,\,(GeV)^{4}\label{51}\end{equation}
which if immediately thermalized would give rise to temperature \begin{equation}
T_{e}\,\,\,\simeq\,\,\,\frac{3.03\times10^{12}}{F_{X}^{1/2}}\,\,\left(\frac{g}{g_{*}}\right)^{1/4}\,\, GeV\label{52}\end{equation}
where $g_{\ast}$ is the total number of species in the thermal bath
and maybe somewhat higher than \textit{g}. We assume that immediately
after the end of inflation there is complete kinetic domination so
that $\rho_{\phi}\,\,\propto\,\,1/a^{6}$. Then we get \begin{equation}
\frac{\rho_{R}}{\rho_{\phi}}\,\,\propto\,\, a^{2}\label{53}\end{equation}
Hence from Eq.\eqref{50} we see that the universe has to expand by
a factor of about $10^{6}$ to $10^{7}$ after the end of inflation
to become radiation dominated and at which stage the temperature which
goes as $T\,\,\propto\,\,1/a$ is given by \begin{equation}
T\,\simeq\,\,\frac{3.03\times10^{5}}{F_{X}^{1/2}}\,\, GeV\label{54}\end{equation}

So we see that radiation domination sets in comfortably before nucleosynthesis.
But the above expression needs some correction to allow for the period
between the end of inflation, when $\rho_{\phi}\,\,\propto\,\,1/a^{2}$,
and complete kinetic domination, i.e., when $\rho_{\phi}\,\,\propto\,\,1/a^{6}$.
Although this will reduce the temperature at the onset of radiation
domination it will still be high enough for a successful nucleosynthesis,
during which a temperature around 1 \textit{MeV} is sufficient.

So far we have examined the dynamics of the inflationary era. We now
try to impose constraints on the model from the matter dominated era
and the present epoch. We have already shown in Eq.\eqref{36} what
the late time energy density of the universe will be. Observations
require that the current magnitude of a cosmological constant be about
$10^{-12}\;\mathrm{\;}(eV)^{4}\mathrm{\;}$. So we must have \begin{equation}
C\,\,\simeq\,\,10^{-48}\,\,(GeV)^{4}\label{55}\end{equation}
Further, since the current dark matter density is about one-third
that of dark energy, one has \begin{equation}
\frac{C}{3}\,\,\simeq\,\,\frac{k}{Aa_{0}^{3}}\label{56}\end{equation}
the subscript $`0'$ signifying the present epoch. Observations tell
us that the fraction of the present total energy density of the universe
contained in radiation is $\left(\Omega_{R}\right)_{0}\,\,\simeq\,\,5\times10^{-5}$
and that contained in dark energy is $\left(\Omega_{DE}\right)_{0}\,\,\simeq\,\,0.73$.
The present radiation density of the universe is thus $\left(\rho_{R}\right)_{0}\,=\,\,\frac{\left(\Omega_{R}\right)_{0}}{\left(\Omega_{DE}\right)_{0}}\,\, C\,\,\simeq\,\,6.94\times10^{-53}\,\,(GeV)^{4}$.
We denote the third term in \eqref{36} as $\rho_{k}$. It is known
that nucleosynthesis occurs at a redshift of $z\,\,\sim\,\,10^{10}$.
We assume that $\rho_{R}$ crosses over $\rho_{k}$ at a redshift
of $z\,\,\sim\,\,10^{12}$. We then get \begin{equation}
{z^{2}\,\,\simeq\,\,\frac{4A^{2}Ba_{0}^{6}}{k^{2}}\,\,\left(\rho_{R}\right)_{0}\,\,=\,\,4\,\,\frac{9}{C^{2}}\, B\,\left(\rho_{R}\right)_{0}}\label{57}\end{equation}
Thus one obtains a lower bound on the parameter \textit{B} given by
\begin{equation}
B\,\,\geq\,\,4\times10^{-22}\,(GeV)^{4}\label{58}\end{equation}
Now using Eq.\eqref{47} we obtain an upper bound on parameter $A$
given by \begin{equation}
A\,\,\leq\,\,\,10^{10}\,\,(GeV)^{-2}\label{59}\end{equation}
Using the limiting values for the parameters it is found that the
cross-over between the dark matter density and $\rho_{k}$ occurs
at a redshift of $z\,\,\sim\,\,10^{9}$, and that between dark matter
and radiation occurs at a redshift of $z\,\,\sim\,\,10^{4}$, i.e.,
at the epoch of matter-radiation equality. We also find that the present
value of $\rho_{k}$ is \begin{equation}
\left(\rho_{k}\right)_{0}\,\,=\,\,\frac{k^{2}}{4A^{2}Ba_{0}^{6}}\,\,\simeq\,\,6.94\times10^{-77}\,\,(GeV)^{4}\label{60)}\end{equation}
and the adiabatic sound speed at the epoch of matter-radiation equality
(at a redshift of about $10^{4}$) is \begin{equation}
\left(c_{s}^{2}\right)_{eq}\,\,=\,\,\dfrac{1}{\dfrac{2ABa_{eq}^{3}}{k}\,\,+\,\,1}\,\,\simeq\,\,\frac{1}{\dfrac{6B}{C\, z_{eq}^{3}}\,\,+\,\,1}\,\,\simeq\,\,4.1\times10^{-16}\label{61}\end{equation}
We can rexpress $w$ from Eq.\eqref{37} in terms of the redshift
$z$ . Since $\rho_{k}$ is negligible compared to the other components,
we have \begin{equation}
w\,\,\approx\,\,\frac{{-\, C}}{{C\,+\,\dfrac{k}{{Aa^{3}}}}}\,\,=\,\,\frac{{-\, C}}{{C\,+\dfrac{k}{Aa_{0}^{3}}\left({z+1}\right)^{3}}}\label{62}\end{equation}
Therafter it is possible to find $dw/dz\ $. Its value at the current
epoch, i.e., at redshift $z=0$ using the above limiting values of
\textit{A} and \textit{B} from Eqs.\eqref{58}, \eqref{59} turns
out to be \begin{equation}
\left(\frac{dw}{dz}\right)_{z=0}\approx2.733\times10^{-28}\label{63}\end{equation}

On the other hand, observations suggest that inflation ended at a
redshift of about $z\,\,\sim\,\,10^{28}$. As we saw in the analysis
on inflationary dynamics, radiation comes to dominate the kinetic
energy density of the scalar field after the universe has expanded
by about $10^{6}$ to $10^{7}$ after the end of inflation. Assuming
that $\rho_{R}$ crosses over $\rho_{k}$ at a redshift of $10^{20}$,
and proceeding as before for obtaining Eq.\eqref{58}, in this case
we obtain an upper bound on the parameter $B$, 
\begin{equation}
B\,\,\le\,\,4\times10^{-6}\,\,\left(GeV\right)^{4}\label{64}
\end{equation}
and then a corresponding lower bound on the parameter \textit{A} (using
\eqref{47}) given by
\begin{equation}
A\,\,\ge\,\,250\,\,\left(GeV\right)^{-2}\label{65}\end{equation}
Using these set of limiting values we find that the cross over between
dark matter and $\rho_{k}$ occurs at a redshift of about $10^{14}$,
whereas that between dark matter and radiation remains the same as
in the earlier case. In this case $(\rho_{k})_{0}$ and $\left(c_{s}^{2}\right)_{eq}$
are given by 
\begin{equation}
\left(\rho_{k}\right)_{0}\simeq6.94\times10^{-93}GeV^{4}\label{66}
\end{equation}
\begin{equation}
\left(c_{s}^{2}\right)_{eq}\simeq4.1\times10^{-32}\label{67}
\end{equation}
If we use the limiting values of \textit{A} and \textit{B} from Eqs.\eqref{64}
and \eqref{65} in the $dw/dz$ relation obtained from Eq.(62), we
get 
\begin{equation}
\left(\frac{dw}{dz}\right)_{z=0}\approx1.281\times10^{-45}\label{68}\end{equation}
One can also estimate the current value of the equation of state parameter
in our model, which using \eqref{37} turns out to be \begin{equation}
w_{0}\simeq\frac{-C}{C+\dfrac{k}{Aa_{0}^{3}}}\simeq\frac{-C}{C+C/3}\simeq-\ 0.75\label{69)}\end{equation}

It should be noted here that the need to determine the value of \textit{k}
explicitly did not arise in our calculations. Its value can be determined
from \eqref{56}, provided we know the values of \textit{A} and \textit{C},
i.e., \textit{k} is not an independent parameter in our model. We
can further find out at what redshift the universe started to accelerate
due to the presence of dark energy. Knowing that for acceleration
to begin we must have $w=-1/3$, from \eqref{62} we find 
\begin{equation}
z_{acc}\approx0.817\label{70}
\end{equation}
Such a value for the redshift is in fact quite compatible with present
observations \cite{melchiorri}. Finally, using Eqs.\eqref{58}, \eqref{59},
\eqref{64} and \eqref{65} in Eq.\eqref{20}, one finds that the parameter
$L$ of our model \eqref{17} is constrained to lie in the
range 
\begin{equation}
10^{-49}\,\le\, L\,\le\,10^{-41}
\end{equation}
and $M$ has to be tuned to satisfy the last relation in Eq.\eqref{20}.
We thus see that for a choice of the parameters $K\sim O(1)$ and
$L$  in the range given above it is possible to have a \textit{k}-essence
model that not only unifies dark matter and dark energy but also produces
inflation in the early universe as well. Note that the requirement of
tuning of one of the parameters, viz., $M$ is to be expected, since this is 
merely a restatement of the fine-tuning
problem associated with the cosmological constant. 
Further, it may be noted that the coincidence problem of the standard
$\Lambda$CDM cosmology is retained at a similar level within the present framework.
In addition to the tuning of the parameter $M$, as in the $\Lambda$CDM model we
have used observations to fix the ratio of $\Omega_m$ and $\Omega_{\Lambda}$
effectively through our Eq.\eqref{56}. Though dark matter and dark energy
are generated within a unified framework in this model, the late time behaviour
is quite akin to that of the standard $\Lambda$CDM model with its coincidence
problem.

\section{Conclusions}

To summarize, we have considered a model of \textit{k}-essence to
study the possibility of producing inflation in the early universe,
and susequently generating both dark matter and dark energy during
later evolution in appropriate order. We have first shown that it
is difficult to unify dark matter and dark energy using purely
kinetic \textit{k}-essence, since the ansatz of a late time energy density
expressed simply as a sum of a cosmological constant and a matter term
leads to a static universe. We have presented an alternative model
including a potential for the scalar field that achieves this unification
and also behaves effectively as purely kinetic \textit{k}-essence
at late times. Our model falls under the class of models 
dubbed \textit{k}-essence
which contain non-canonical kinetic terms. We have shown that our
model generates inflation in the early universe that reproduces the
basic features of the standard chaotic inflation model involving a
quadratic potential. At the end of inflation when the potential in
our model becomes negligible in comparison to the kinetic component
we were able to approximate the model as purely kinetic \textit{k}-essence.
The expression for the energy density in terms of the scale factor
$a$ and also for that of adiabatic sound speed were obtained. We
found that the resultant energy density contained terms that achieved
the unification of dark matter and dark energy. The adiabatic sound
speed came out to be close to zero when calculated at the epoch of
matter-radiation equality, thus posing no problems for structure formation,
since the sound speed decreases further as the scale factor increases.
Current observations quite strongly favour a cosmological constant
as the source of dark energy. Our model reproduces a cosmological
constant at late times. The value of the current equation of state
parameter, and the red-shift at which the transition to the accelerated
phase occurs, that we estimated, lies within observational bounds.

We considered a general form for the \textit{k}-essence Lagrangian
containing a non-canonical kinetic term. We then used observational
constraints ranging from the inflationary era to the subsequent matter
and radiation dominated eras and the present accelerated phase as
well to impose a set of bounds on the model parameters. In this way
we could provide an estimate of the relative strengths of the various
terms of our model Lagrangian. It should be pointed out that the form
of the potential chosen for the model, though widely used for its
simplicity, is not very realistic and only serves to highlight the
features of the model during the inflationary era. Recent WMAP data
analysis \cite{destri} suggest that the best fit potential for inflation
is a trinomial potential and further study of our model could be made
by using such a potential. Moreover, it would be interesting to investigate
the relation of our model to the dynamics of another widely used class
of \textit{k}-essence models where the Lagrangian is taken to be of
the type ${\cal L}=F(X)V(\phi)$. Finally, since the consideration
of non-canonical scalar field kinetics in cosmology was originally
motivated by the Born-Infeld \cite{born-infeld} action of string
theory, and there have been many more recent string theoretic inputs
in cosmology such as the idea of the landscape \cite{landscape},
it should be worthwhile to the explore the possible origin of generalized
non-canonical actions such as ours in the low energy limit of specific
string theoretic models.

\section*{Acknowledgements}

A.S.M. would like to acknowledge support from a project funded by DST, India.

\end{document}